\documentclass[reprint,aps,prl]{revtex4-1}

\usepackage{amsmath,amssymb,amsfonts} 
\usepackage{graphicx}
\usepackage{dcolumn}
\usepackage{bm}
\usepackage[bookmarks=true,colorlinks=true,citecolor=myblue,linkcolor=myblue,filecolor=myblue,urlcolor=myblue]{hyperref} 
\usepackage{braket} 
\usepackage{tikz} 
\usepackage{comment} 
\usepackage[normalem]{ulem} 
\usepackage{booktabs}
\usepackage{orcidlink}

\DeclareMathOperator{\Tr}{Tr}

\definecolor{myblue}{RGB}{0, 170, 212}
\definecolor{mygreen}{RGB}{0,220,0}

\begin{document}

\title{Truncation error in series expansions linking microcanonical and canonical ensembles}

\author{Atsushi Iwaki\,\orcidlink{0000-0002-5345-8045}}
\email{atsushi-iwaki@phys.s.u-tokyo.ac.jp}
\affiliation{Department of Physics, University of Tokyo, 7-3-1 Hongo, Bunkyo-ku, Tokyo 113-0033, Japan}

\date{\today}

\begin{abstract}
In the microcanonical thermal pure quantum (mTPQ) method, 
the canonical ensemble is derived using Taylor series expansions. 
We prove that 
the truncation error decreases exponentially with system size 
when the effective temperature of the mTPQ state is smaller than the target temperature, 
and otherwise, the error remains constant. 
We also show the discipline to set the mTPQ parameter  
by considering the trade-off between the error and the numerical cost.
\end{abstract}

\maketitle

\emph{\textbf{Introduction.---}}
In various fields of physics, 
it is essential to calculate the thermal expectation value of the canonical ensemble, 
expressed as $\langle O \rangle_\beta = \Tr[\rho_\beta \Hat{O}]$ 
where $\rho_\beta = e^{-\beta\Hat{H}} / \Tr[e^{-\beta\Hat{H}}]$ 
with the Hamiltonian $\Hat{H}$ and the inverse temperature $\beta$. 
One of the major numerical approaches for this is random sampling; 
prepare an initial random state $\ket{0}$ with $\overline{\ket{0}\bra{0}} \propto \Hat{I}$ 
and perform imaginary-time evolution as $\ket{\beta} = e^{-\beta\Hat{H}/2} \ket{0}$. 
Here, $\overline{\cdots}$ denotes random averages. 
By repeating this process a sufficient number of times
and taking appropriate sample averages, 
exact physical quantities can be acquired \cite{iwaki2022}. 
In the microcanonical thermal pure quantum (mTPQ) method \cite{sugiura2012}, 
instead of imaginary-time evolution, 
we iteratively apply the Hamiltonian as $\ket{k} = (l-\Hat{h})^k \ket{0}$ 
where $\Hat{h} = \Hat{H}/N$ is the Hamiltonian normalized by the system size
and $l$ is a real parameter larger than the maximum eigenvalue of $\Hat{h}$. 
To determine physical quantities associated with $\ket{\beta}$ using those of $\ket{k}$, 
a formula presented in Eq.(13) of Ref.~\cite{sugiura2013} is commonly used: 
\begin{multline}
    e^{N\beta l} \overline{\braket{\beta|\Hat{O}|\beta}} = \\
    \sum_{k=0}^{\infty} \frac{(N\beta)^{2k}}{(2k)!} \overline{\braket{k|\Hat{O}|k}} + 
    \sum_{k=0}^{\infty} \frac{(N\beta)^{2k+1}}{(2k+1)!} \overline{\braket{k|\Hat{O}|k+1}}. 
    \label{eq:sugiura}
\end{multline}
This formula is proven using the Taylor series expansion shown below: 
\begin{align}
    \Tr [ e^{N \beta (l-\Hat{h})} \Hat{O} ] 
    &= \sum_{k=0}^{\infty} \frac{(N\beta)^k}{k!} 
    \Tr [ (l-\Hat{h})^k \Hat{O} ]. 
    \label{eq:taylor}
\end{align}
The mTPQ method has several advantages: 
the absence of Trotter error which frequently occurs in other methods 
and the ability to access arbitrary temperatures by storing $\braket{k|\Hat{O}|k}$ and $\braket{k|\Hat{O}|k+1}$. 
Numerical packages like H$\Phi$ \cite{ido2024} implement the mTPQ method. 
It is also possible to utilize a random matrix product state as the initial state $\ket{0}$ 
instead of a Haar random state \cite{iwaki2021}.

The thermal state $\ket{k}$ is associated with the smoothed microcanonical ensemble 
expressed as $\rho_k = (l-\Hat{h})^k / \Tr[(l-\Hat{h})^k]$, 
which is one of the squeezed ensembles \cite{yoneta2019}. 
The effective inverse temperature of $\rho_k$ is given by 
\begin{align}
    \beta_k = \frac{k}{N} \frac{1}{l-u_k}, 
    \label{eq:beta_k}
\end{align}
where $u_k = \langle h \rangle_k = \Tr[\rho_k \Hat{h}]$. 
The parameter $k$ needs to scale linearly with the system size 
to reach a temperature independent of the system size. 
We note that the effective inverse temperature is represented by $\beta_{2k}$ in the mTPQ method, 
as $\ket{k}$ strictly corresponds to $\rho_{2k}$. 

In practical simulations, 
the series expansion must be truncated at a finite $k$. 
Despite the lack of theoretical understanding of the optimal $k$, 
empirical data indicates that the series approximation performs well 
when the effective temperature $\beta_k$ is lower than the target temperature $\beta$. 
In this work, we present a rigorous explanation for this empirical finding. 
We also investigate the relationship between the truncation error 
and the characteristic parameter $l$ in the mTPQ method. 
We note that 
similar analyses have been conducted in the Appendices of Refs.~\cite{sugiura2013, yamaji2018}, 
but our calculations yield more rigorous results. 

\emph{\textbf{Main results.---}}
We assume that the Taylor series expansion in Eq.~\eqref{eq:taylor} is truncated at $k$. 
By Taylor's theorem, 
the remainder term can be represented as 
\begin{align}
    R_k 
    &= \frac{(N\beta)^{k+1}}{k!} \int_{0}^{1} dt~ t^k 
    \Tr[e^{N\beta(l-\Hat{h})(1-t)}(l-\Hat{h})^{k+1}\Hat{O}] \notag \\
    &= \frac{(N\beta)^{k+1}}{k!} \int_{0}^{1} dt~ t^k \int du~
    e^{Ns_\mathrm{th}(u)} \notag \\
    &\quad\quad\quad\quad\quad\quad 
    \times e^{N\beta(l-u)(1-t)} (l-u)^{k+1} \langle O \rangle_u, 
\end{align}
where $s_\mathrm{th}(u)$ is the thermal entropy density 
and $\langle O \rangle_u$ is the microcanonical expectation value 
at energy density $u$. 
Since $k$ scales linearly with the system size as $k = N \kappa$, 
we can apply Laplace's method to evaluate the remainder: 
\begin{gather}
    R_k
    = \frac{(N\beta)^{k+1}}{k!} \int_{0}^{1} dt \int du~ (l-u) \langle O \rangle_u 
    ~e^{Ng(t, u)}, \\
    g(t,u) = \kappa\log [t(l-u)] + s_\mathrm{th}(u) + \beta(l-u)(1-t). 
\end{gather}
To find the maximum point of $g(t, u)$, we solve the following system of equations: 
\begin{gather}
    \frac{\partial}{\partial t} g = \frac{\kappa}{t} - \beta(l-u) = 0, \\
    \frac{\partial}{\partial u} g = \beta_\mathrm{th} (u) - \beta(1-t) - \frac{\kappa}{l-u} = 0. 
\end{gather}
Here, $\beta_\mathrm{th}(u) = s_\mathrm{th}'(u)$ is the inverse temperature 
as a function of energy density. 
Let $(\Tilde{t}, \Tilde{u})$ denote the maximum point. 
Upon solving for $\Tilde{u}$, we obtain $\beta_\mathrm{th}(\Tilde{u}) = \beta$, 
suggesting that $\Tilde{u}$ is the canonical value at inverse temperature $\beta$: 
\begin{align}
    \Tilde{u} = u_\beta = \langle h \rangle_\beta. 
\end{align}
Then, we calculate $\Tilde{t}$ as 
\begin{align}
    \Tilde{t} = \frac{\kappa}{\beta(l - u_\beta)}. 
\end{align}
We need to verify whether $\Tilde{t}$ is within the integral range $[0, 1]$. 
If $\Tilde{t} \le 1$, we find 
\begin{align}
    \beta_k (l-u_k) \le \beta(l-u_\beta). 
    \label{eq:ineq}
\end{align}
We now examine the behavior of $\beta(l-u_\beta)$ as a function of $\beta$: 
\begin{align}
    \frac{d}{d\beta} \left[ \beta(l-u_\beta) \right] = l-u_\beta + \frac{c_\beta}{\beta} > 0. 
\end{align}
Here, $c_\beta = - \beta^2 ~ du_\beta / d\beta$ is the specific heat density. 
Because $\beta(l-u_\beta)$ is a monotonically increasing function, 
the inequality \eqref{eq:ineq} is equivalent to $\beta_k \le \beta$. 

When $\beta_k \le \beta$, 
the maximum point $(\Tilde{t}, \Tilde{u})$ is included in the integral range. 
This allows for the evaluation of the remainder as 
\begin{align}
    R_k \sim \langle O \rangle_\beta 
    \exp N \beta [l - f_\mathrm{th}(\beta)], 
\end{align}
by focusing on exponentially large factors, 
where $f_\mathrm{th} (\beta) = u_\beta - s_\mathrm{th}(u_\beta) / \beta$ 
is the thermodynamic free energy density. 
The relative error due to the truncation is then estimated as 
\begin{align}
    \epsilon_k = \frac{R_k}{\Tr[e^{N\beta(l-\Hat{h})}\Hat{O}]} \sim 1. 
\end{align}
This indicates that there remains a substantial truncation error. 
Namely, the series approximation does not work when $\beta_k \leq \beta$. 

On the other hand, when $\beta_k > \beta$, 
contributions near $t=1$ dominate the integral in the thermodynamic limit. 
Therefore, the remainder can be approximated as 
\begin{align}
    R_k 
    &\simeq \frac{(N\beta)^{k+1}}{k!} \int du~
    e^{Ns_\mathrm{th}(u)} (l-u)^{k+1} \langle O \rangle_k. 
\end{align}
This integral reaches its maximum at $u = u_k$, 
allowing us to apply Laplace's method once more: 
\begin{align}
    R_k
    \sim \langle O \rangle_k \exp N [\kappa + s_\mathrm{th}(u_k) + \kappa\log\beta - \kappa\log\beta_k]. 
\end{align}
The relative error is then evaluated as 
\begin{align}
    \epsilon_k 
    &\sim \frac{\langle O \rangle_k}{\langle O \rangle_\beta} \exp N[h(\beta) - h(\beta_k)], 
\end{align}
where $h(\beta) = \beta f_\mathrm{th}(\beta) + \kappa\log\beta - l\beta$. 
Let us take the derivative of $h(\beta)$ as 
\begin{align}
    h'(\beta) = \frac{\beta_k(l-u_k) - \beta(l-u_\beta)}{\beta}, 
\end{align}
where we find $\beta_k(l-u_k) > \beta(l-u_\beta)$ for $\beta_k > \beta$, 
thus $h(\beta)$ monotonically increases and $h(\beta_k) > h(\beta)$. 
Consequently, the truncation error decreases exponentially with the system size. 
For small values of $\Delta\beta_k = \beta_k - \beta$, 
the error is approximated as 
\begin{align}
    \epsilon_k 
    &\sim \exp \left[ -N \frac{(\Delta\beta_k)^2}{\beta} \left( l - u_\beta + \frac{c_\beta}{\beta} \right) \right]. 
    \label{eq:eps_l}
\end{align}
The parameter $l$ controls the degree of this exponential decay of the error. 

\emph{\textbf{Discussion.---}}
We have demonstrated that the relative truncation error of the series expansion in Eq.~\eqref{eq:taylor} 
is exponentially small if $\beta_k > \beta$, 
whereas remains constant if $\beta_k \le \beta$. 
Notice that in proving this result, 
we have used some equations 
that are exact only in the thermodynamic limit, 
thus, the conclusions are approximately valid for finite $N$.

Finally, we discuss how to choose the parameter $l$ in the mTPQ method. 
A larger value of $l$ effectively suppresses the truncation error as shown in Eq.~\eqref{eq:eps_l}. 
However, a larger $k$ is then required to reach the same temperature as indicated in Eq.~\eqref{eq:beta_k}. 
Therefore, $l$ should be determined, 
considering the trade-off between the reduction of the truncation error 
and the computational cost associated with $k$. 

\emph{\textbf{Acknowledgments.---}}
We thank Chisa Hotta for fruitful discussions.  
This work was supported by the Center of Innovation for Sustainable Quantum AI, 
JST Grant Number JPMJPF2221. 

\nocite{apsrev41Control}
\bibliographystyle{apsrev4-1}
\bibliography{references}

\end{document}